\documentclass[letterpaper, 10 pt, conference]{ieeeconf}  % Comment this line out if you need a4paper
\IEEEoverridecommandlockouts                              % This command is only needed if 
\usepackage[pdftex,pdftitle={Quickest Change Detection Approach to Optimal Control in Markov Decision Processes with Model Changes}]{hyperref}
\hypersetup{colorlinks,linkcolor={green!50!black},citecolor={green!50!black},urlcolor={blue!80!black}}
\makeatletter \let\NAT@parse\undefined \makeatother
\usepackage[sort,compress]{cite}

\usepackage{graphics} % for pdf, bitmapped graphics files
\usepackage{epsfig} % for postscript graphics files
\usepackage{mathptmx} % assumes new font selection scheme installed
\usepackage{times} % assumes new font selection scheme installed
\usepackage{amsmath} % assumes amsmath package installed
\usepackage{amssymb}  % assumes amsmath package installed

\usepackage{balance}

\usepackage[utf8]{inputenc} % allow utf-8 input
\usepackage{hyperref}       % hyperlinks
\usepackage{url}            % simple URL typesetting
\usepackage{booktabs}       % professional-quality tables
\usepackage{amsfonts}       % blackboard math symbols
\usepackage{nicefrac}       % compact symbols for 1/2, etc.
\usepackage{microtype}      % microtypography
\usepackage{subfig}
\usepackage{graphicx}
\usepackage{algpseudocode}

\makeatletter
\newcommand{\algrule}[1][.5pt]{\par\vskip.5\baselineskip\hrule height #1\par\vskip.5\baselineskip}
\makeatother

\usepackage[svgnames]{xcolor} \definecolor{DarkGreen}{rgb}{0,0.5,0}
\definecolor{DarkRed}{rgb}{0.75,0,0}

\newtheorem{theorem}{Theorem}%[section]

\newtheorem{lemma}[theorem]{Lemma}

\newtheorem{algorithm}{Algorithm}[section]

\newcommand{\Prob}{\mathbb{P}}
\newcommand{\Expect}{\mathbb{E}}

\long\def\symbolfootnote[#1]#2{\begingroup%
\def\thefootnote{\fnsymbol{footnote}}\footnote[#1]{#2}\endgroup}

\usepackage{color}
\definecolor{DarkRed}{rgb}{0.75,0,0}
\definecolor{DarkGreen}{rgb}{0,0.5,0}
\definecolor{DarkBlue}{rgb}{0,0,0.5}
\definecolor{DarkPurple}{rgb}{0.5,0,0.5}
\title{\LARGE \bf Quickest Change Detection Approach to Optimal Control in Markov Decision Processes with Model Changes}

\author{Taposh Banerjee$^{1}$, Miao Liu$^{2}$ and Jonathan P. How$^{3}$% <-this % stops a space
	\thanks{$^{1}$Taposh Banerjee is with the School of Engineering and Applied Sciences, Harvard University,
		Cambridge, MA.       
		{\tt\footnotesize tbanerjee@seas.harvard.edu}}%
	\thanks{$^{2}$Miao Liu is with IBM T. J. Watson Research Center, Yorktown Heights, New York. 
		{\tt\footnotesize miao.liu1@ibm.com}}%
	\thanks{$^{3}$Jonathan P.\ How is with the Laboratory for Information and Decision Systems (LIDS),
		Massachusetts Institute of Technology, Cambridge, MA.       
		{\tt\footnotesize jhow@mit.edu}}%
	\thanks{Work completed while the first two authors were at LIDS, MIT.}
}

\begin{document}
	
	\maketitle
	\thispagestyle{empty}
	\pagestyle{empty}

	\begin{abstract}
		%The problem of optimal control of a Markov decision process (MDP) with model changes is considered. 
		%At various unknown points in time the model of the MDP undergoes changes. The objective 
		%is to maximize long term discounted rewards in such a non-stationary environment. 
		Optimal control in non-stationary Markov decision processes (MDP) is a challenging problem. The aim in such a control 
		problem is to maximize the long-term discounted reward when the transition dynamics or the reward function can change over time. 
		When a prior knowledge of change statistics is available, the standard
		Bayesian approach to this problem is to reformulate it as a partially observable MDP (POMDP)
		and solve it using approximate POMDP solvers, which are typically computationally demanding. 
		In this paper, the problem is analyzed through the viewpoint of quickest change detection (QCD), a set of tools 
		for detecting a change in the distribution of a sequence of random variables. 
		Current methods applying QCD to such problems only passively detect changes by following prescribed policies, without optimizing the choice of actions for long term performance. We demonstrate that ignoring the reward-detection trade-off can cause a significant loss in long term rewards, and propose a two threshold switching strategy to solve the issue. 
		A non-Bayesian problem formulation is also proposed for scenarios where a Bayesian formulation cannot be defined. 
		The performance of the proposed two threshold strategy is examined through numerical analysis on a non-stationary MDP task, and the strategy outperforms the state-of-the-art QCD methods in both Bayesian and non-Bayesian settings. 
	\end{abstract}
	
	\section{Introduction and Motivation}
	In their life time, autonomous agents may be required to deal with non-stationary stochastic environments, where parameters of the world (dynamics, cost and goals) change over time (for the same states spaces). For example, in robotic navigation, %in human-aware UAV path planning in urban environments~\cite{allamaraju2014human}, 
	a path planner might need to accommodate dynamically changing friction coefficients of road surfaces or wind conditions to ensure safety. In inventory control, an order management agent might need to account for the time varying distribution of demands for goods and services when determining the order at each decision period. %In robot-soccer domains, defenders need to account for the uncertainty and change of opponents' attacking strategies to achieve efficient defense. 
	A service robot may have to adapt to different personality or behavior of the person it is assisting. In a multi-agent system, 
	an agent may have to adapt to changing strategies of the opponent or other players \cite{hernandez2016identifying}. 
	In order to maximize long term rewards, it is paramount that the agents are able to quickly detect the changes in the environment and adapt their policies accordingly. %in human population density to ensure safety; in robot-soccer domains, defenders need to account for the uncertainty and change of opponents' attacking strategies to achieve efficient defense.\mX{Another example is inventory management, where the distribution of demands for goods and services may vary over time.~\cite{A citation}} 
	
	When the environment is stationary, many of these problems for autonomous agents, path planning or robot control can be formulated as an MDP, for which there is a rich literature \cite{bertsekas1995dynamic}.%, \cite{kaelbling1996reinforcement}. 
	When the environment is non-stationary, there is a limited understanding on the best strategy to employ.
	In this paper, we are interested in the case where the non-stationary process consists of several stationary processes. Each stationary process corresponds to an MDP, and the non-stationary nature is captured by changes in the 
	transition and/or reward structure of the MDPs. 
	That is, we consider the whole operation in a non-stationary environment as a global task, which can be decomposed into several stationary subtasks. This is a general case often encountered in real world applications, such as robotics, inventory controls, etc.
	For example, in inventory control, it is reasonable to assume that the demand rate would be steady for some time, before changing to another value. For path planning, one can assume that the road surfaces or wind conditions are steady for a while before changing, etc. 
	
	Solving dynamic decision making in non-stationary environments optimally is an extremely hard problem. Although it is extensively studied in the literature, due to the analytical intractability of the problem, however, existing works aim towards developing approximate solutions. %We discuss some of these results in this section. 
	For example, in some papers, the problem is reformulated as a Partially Observable MDP (POMDP), and approximate POMDP solvers
	are used to obtain a solution; e.g., see \cite{hadoux2014solving} and \cite{dayan1996exploration}. In others, estimates are maintained of the current MDP parameters, and next state or reward predictors are used to assess whether the parameters of the active MDP has changed; see e.g. \cite{doya2002multiple} and \cite{da2006dealing}. In \cite{choi2000hidden} and 
	\cite{chades2012momdps}, new models such as hidden mode MDPs (hmMDPs) and mixed observability MDPs (MOMDPs) are proposed to capture the transition between different MDPs, and approximate solutions are obtained. Although these approximate solutions are promising tools for solving non-stationary MDPs, they are restricted to problems where the model sizes are known a priori, and they might still suffer from high computational overhead. The approach of multitask or transfer learning has also been used to study these problems. 
	However, in these approaches, tasks are usually assumed to be well separated and given before learning and planning take place. There is no explicit concept of automatic change detection in these problems. 
	Problems of this nature are also explored in the adaptive control literature. For example, in \cite{borkar1979adaptive} and \cite{borkar1982identification}, 
	the authors study optimal control of an MDP with unknown parameters in the transition kernel. Again, there is no concept of 
	a change in this problem. %For a stochastic approximation based approach to decision making in non-stationary environments, see 
	%the recent work in \cite{SinghBanerjee2017}. 
	
	In this paper, we approach the problem of optimal control in MDPs with model changes 
	through the viewpoint of classical quickest change detection (QCD) 
	\cite{veeravalli2013quickest}. 
	The changes in properties of MDPs (transition or reward) cause a change in the law of the state-action sequence. 
	Tools and ideas from the QCD literature can be used to detect these changes in law. QCD algorithms are 
	scalable, and are optimized for quickest detection. In spite of these properties, these tools 
	are either never used, or are not fully exploited in science and engineering literature, especially in the problem of interest in this paper. 
	For example, sequential detection approaches are applied in \cite{allamaraju2014human} and \cite{hadoux2014sequential}, where
	the optimal policy for each MDP is executed, and a change detection algorithm 
	is employed to detect model changes. We will show in this paper that such a naive approach may lead 
	to significant loss in performance. The key idea is that what is optimal for optimizing rewards may not 
	necessarily be optimal for quickest detection of model changes. Thus, there is a fundamental reward-detection 
	trade-off to be exploited. The purpose of this paper is to articulate this trade-off in a mathematically precise manner, 
	and to propose solutions that we claim best exploit this trade-off. 
	Specifically, we propose a computationally efficient, simple two-threshold strategy to quickly detect model 
	changes, without significant loss in rewards. We show that such an approach is superior to 
	the existing methods in the literature. Specifically, we show that the two-threshold strategy leads to 
	better performance as compared to the single threshold classical QCD tests, as employed in \cite{hadoux2014sequential}. 
	Note that it is shown in \cite{hadoux2014sequential} that a QCD approach is better (with respect to the Bayesian problem we use in this paper; see \eqref{eq:Prob} below) than other approaches in the literature, e.g., \cite{da2006dealing}. For benchmarking purposes, we also compare with the MOMDP solution \cite{chades2012momdps} and a random action strategy. 
	
\iffalse
	The paper is organized as follows. In Section~\ref{sec:ProbForm}, we discuss the problem formulations used
	in the paper. We consider both Bayesian and non-Bayesian formulations. In the Bayesian setting, we show how this problem can be reformulated as a POMDP, solving which is computationally prohibitive. In Section~\ref{sec:QCD},
	we provide a brief overview of the QCD literature relevant to this paper. The QCD approach allows the study  of both Bayesian and non-Bayesian formulations in a common framework. In Section~\ref{sec:QCD_Local},
	we show a naive way to apply QCD algorithms to the problem and discuss its implications. Specifically, 
	we discuss a reward-detection trade-off that is not exploited by the naive approach. In Section~\ref{sec:QCD_Twothres}, we discuss a two-threshold strategy 
	that exploits the reward-detection trade-off and outperforms the naive approach. Finally, in Section~\ref{sec:numerical}
	we provide numerical results for an inventory control problem to justify our claims. Throughout the paper we mostly discuss the case where there is a single model switch from one MDP to another, but Section~\ref{sec:multModel_multCP} shows how these techniques can be extended to account for multiple change points, and also multiple models, potentially an infinite number of them. 
\fi

\vspace{-0.2cm}
	\section{Problem Formulation}\label{sec:ProbForm}
	
%\vspace{-0.1cm}

	Assuming we have a family of MDPs $\{\mathbf{M}_\theta\}$, where $\theta$ takes value in some index set $\Theta$, which could be finite or infinite. For each $\theta$, an MDP is defined by a four tuple: $\mathbf{M}_\theta=(\mathcal{S}, \mathcal{A}, \mathcal{T}_\theta, \mathcal{R}_\theta )$, where $\mathcal{S}$ and $\mathcal{A}$ are respectively the state space and the 
	action space common to all MDPs, $\mathcal{T}_\theta$ is the transition kernel, and $\mathcal{R}_\theta$ is the reward function. 
	A decision maker observes a sequence of states $\{s_k\}_{k \geq 0}$, $s_k \in \mathcal{S}, \forall k$, and for each observed state $s_k$, it chooses an action $a_k \in \mathcal{A}$. For each pair $(s_k, a_k)$, the next state $s_{k+1}$ takes values according
	to the law dictated by the kernel: $\forall k$
	\begin{equation*}%\vspace{-0.3cm}
		\begin{split}
			\mathcal{T}_\theta(s, a, s') := &\Prob(s_{k+1} = s' \;  | \; a_k = a, s_k = s), \; s,s' \in \mathcal{S}, a \in \mathcal{A}.
			%&\quad \quad \mbox{ for } s,s' \in \mathcal{S}, a \in \mathcal{A}, \forall k.
		\end{split}
	\end{equation*}
	The reward obtained by choosing action $a_k$ after observing $s_k$ is given by $\mathcal{R}_\theta(s_k, a_k)$.
	
	We operate in a non-stationary environment, which means that the transition structure or the reward structure can change over time. 
	Specifically, at time say $\gamma_1$, 
	the parameter $\theta$ of the MDP changes from $\theta=\theta_0$ to say $\theta =\theta_1$. At a later time, say $\gamma_2$, the 
	parameter changes from $\theta=\theta_1$ to $\theta = \theta_2$, and so on. %Note that a change in the value of the 
	%parameter $\theta$ indicates the change of MDP models $\mathbf{M}_\theta$. 
	To be precise, denote the non-stationary dynamics for $k \geq 0$ as
	\begin{equation}
		\begin{split}
			\Prob(s_{k+1} = s'  \;   | \; a_k = a,& s_k = s) \\
			&\!\!\!\!\!\!\!\!\!\!=
			\left\{ 
			\begin{array}{l l}
				\mathcal{T}_{\theta_0}(s, a, s'), \text{\;\;for $k < \gamma_1$}\\
				\mathcal{T}_{\theta_1} (s, a, s'), \text{\;\;for $\gamma_1\!\leq\!k < \gamma_2$}
			\end{array} \right.\cr
			\mbox{ Reward for } (s_k, a_k) &:= R_k (s_k, a_k) \\
			&\!\!\!\!\!\!\!\!\!\!=
			\left\{ 
			\begin{array}{l l}
				\mathcal{R}_{\theta_0}(s_k, a_k), \hspace{0.0cm}\mbox{ for } k < \gamma_1\\
				\mathcal{R}_{\theta_1} (s_k, a_k), \hspace{0.0cm} \mbox{ for } \gamma_1 \leq k < \gamma_2. 
			\end{array} \right.
		\end{split}
	\end{equation}

	For simplicity of exposition, for now we restrict our attention to the case where there is only one change point $\gamma$, 
	and there are only two models, a pre-change 
	model $\mathbf{M}_0 = (\mathcal{S}, \mathcal{A}, \mathcal{T}_0, \mathcal{R}_0 )$ and a post-change 
	model $\mathbf{M}_1 = (\mathcal{S}, \mathcal{A}, \mathcal{T}_1, \mathcal{R}_1 )$. 
	In Section~\ref{sec:multModel_multCP}, we discuss extensions of the ideas developed for the two model case, 
	to the case where there are more than two models (possibly infinite), and possibly more than one change point. 
	
	We consider two different problem formulations: Bayesian and Non-Bayesian. 
	
	\subsection{Bayesian Formulation}
	Suppose we have a prior on the single change point $\gamma=\gamma_1$, and we have chosen a policy $\Pi=(\pi_1, \pi_2, \cdots)$ that maps the states $\{s_k\}$ to actions $\{a_k\}$. The MDPs $\mathbf{M}_0$ and $\mathbf{M}_1$, the policy $\Pi$, and the prior on the change 
	point, together induce a joint distribution on the product space of state spaces and actions which we denote by $\Prob$, and 
	the corresponding expectation by $\Expect$. The objective is to choose the policy $\Pi$ so as to maximize a long term reward. 
	\begin{equation}\label{eq:Prob}
		\max_\Pi \; \Expect\left[  \sum \nolimits_{k=0}^\infty \beta^k R_k (s_k, a_k)\right],
	\end{equation}
	where $\beta \in [0,1)$ is a discount factor. %When $\beta=1$, we assume that there is a terminal state where the MDP truncates
	%and hence the total cost if finite. 
	Note that the change point $\gamma$ is unobservable. As a result, a Markov policy may not be optimal, and 
	the policy $\Pi$ above has to use the past history to choose the optimal action. To accommodate model changes,  problem~\eqref{eq:Prob} can be reformulated as a Partially Observable MDP (POMDP) (see Lemma~\ref{lem:POMDP} below), and can be solved using approximate POMDP solvers. However, it is computationally expensive to solve this problem numerically, 
	especially when the number of possible models is large. 
	We thus use other approaches to obtain approximate solutions. 
	The vehicle we have chosen to obtain the approximate solutions is the theory of QCD. 
	We show in Section~\ref{sec:numerical} that our proposed solutions performs as well as the solution obtained from 
	standard approximate POMDP solvers in the literature, while at the same time it incurs significantly less computational cost, 
	online and off-line. 
	
	\begin{lemma}\label{lem:POMDP}
		The non-stationary MDP problem~\eqref{eq:Prob} is equivalent to a POMDP problem with the model specified as  $\mathbf{M}_{POMDP}(\mathcal{X}, \mathcal{A}, \mathcal{O}, \widetilde{\mathcal{T}}, \Omega,\widetilde{\mathcal{R}})$, where $\mathcal{X}=\mathcal{S}\times\Theta$ is the POMDP state space, which concatenates the fully observable MDP state space $\mathcal{S}$ and the unobservable model index space $\Theta$; $\mathcal{O}$ is the set of observations, which is equivalent to the set of MDP states; $\widetilde{\mathcal{T}}:X\times A\times X\rightarrow [0,1]$ is the POMDP state transition function; $\Omega:\mathcal{X}\times \mathcal{A}\times \mathcal{O}\rightarrow[0, 1]$ is the observation function; $\widetilde{\mathcal{R}}:\mathcal{X}\times \mathcal{A}\rightarrow\mathbb{R}$ is the reward function with
		\begin{equation}\label{eq:recomputed-value}
			\begin{split}
				\widetilde{\mathcal{T}} \big(x' = (s',\theta')  | x =(s,\theta),a\big) &= \mathcal{T}_{\theta'}(s, a, s') \mathcal{F}(\theta'|\theta)\mathbb{I}(\theta'\neq\theta) 
				\\
				& +  \mathcal{T}_\theta(s,a, s')\mathcal{F}(\theta'|\theta)\mathbb{I}(\theta'=\theta)
				\\
				\omega(o|x',a) \; =& \; \mathbb{I}(o=s')
				\\
				\tilde{\mathcal{R}}(x,a) \; =& \; \mathcal{R}_{\theta}(s,a)
			\end{split}
		\end{equation}
		where $\mathbb{I}(\cdot)$ is an indicator function, and $\mathcal{F}$ is a model transition matrix. Given the prior over a change point to be geometric distribution, i.e.,$\gamma\sim Geom(\lambda) $, it can be shown that
		\begin{equation}\label{eq:beta-backward}
			\mathcal{F}(\theta'|\theta) = \left\{ 
			\begin{array}{l l}
				\lambda\mathbb{I}(\theta'\neq\theta)\!\!+\!\!(1-\lambda)\mathbb{I}(\theta'=\theta)\text{\;if $\theta=0$}\\
				\mathbb{I}(\theta'\neq\theta) \text{\;\;\;\;\;\;\;\;\;\;\;\;\;\;\;\;\;\;\;\;\;\;\;\;\;\;\;\;\;\;if $\theta=1$}
			\end{array} \right. .
		\end{equation}
		
	\end{lemma} 
	
	\vspace*{.1in}
	By converting the original non-stationary MDP problem into a POMDP problem, one can apply any existing POMDP solvers to obtain an approximate POMDP policy, which maps the belief state $b(x)$ to an action $a$. Considering the fact that here the state space can be factorized into an observable part and an unobservable part, this POMDP model can be treated as a mixed observable MDP (MOMDP). In an MOMDP, the belief state becomes a union of $|S|$ disjoint $|\Theta|$-dimension subspaces, and all the operations are performed in the lower dimension, hence is more time efficient to solve a MOMDP than to solve a POMDP.

	\subsection{Non-Bayesian Formulation}
	In practice, a prior on the change point is not often known. Thus, a Bayesian formulation cannot be defined. 
	If the change point is treated as an unknown constant, there are infinite possible laws on the state-action joint spaces, one for each possible change point. In statistics, 
	there are two ways to study non-Bayesian problems: minimax or maxmin and Neyman-Pearson type criteria. 
	
	For minimax or maxmin criterion we may consider the following problem:
	\[
	\max_\Pi \; \min_\gamma \Expect_\gamma \left[ \sum \nolimits_{k=0}^\infty \beta^k R_k (s_k, a_k)\right],
	\]
	where $\Expect_\gamma$ is the expectation with respect to the probability measure when change occurs at time $\gamma$. 
	However, there are two major issues with such an approach in practice. First, a minimax approach may lead to 
	a conservative or pessimistic approach to policy design. Secondly, it is not always straightforward to compute such costs for 
	comparing various competing algorithms.

	Since the problem under consideration is about identifying the correct regime, we also consider a Neyman-Pearson type of criterion: trying to maximize reward in one regime subject to a constraint on the performance under alternative regime. 
	We take this approach (in addition to the Bayesian formulation) in our paper and consider the following problem. 
	\begin{equation}\label{eq:ProbNonBayes}
		\begin{split}
			\max_\Pi \; \quad \Expect_1 & \left[ \sum \nolimits_{k=0}^\infty \beta^k R_k (s_k, a_k)\right] ,\\
			\mbox{ subj. to } \quad \Expect_\infty &\left[\sum \nolimits_{k=0}^\infty \beta^k R_k (s_k, a_k)\right] \geq \alpha. 
		\end{split}
	\end{equation}
	Thus, the objective is to optimize rewards when the change occurs at time $1$, subject to a constraint on the performance
	when the change never occurs. In sequential analysis literature, one often uses a minimax objective rather than the average under $\Expect_1$. As mentioned earlier, this leads to a pessimistic viewpoint. Also, for many sequential detection algorithms 
	in the literature, the worst case is often achieved for $\gamma=1$. Thus, $\Expect_1 \left[ \sum_{k=0}^\infty \beta^k R_k (s_k, a_k)\right]$ can also be seen as a proxy for the minimax or maxmin reward. 
	
	Another advantage of the Neyman-Pearson type criterion is that it maintains a balance in reward structure between the pre- and 
	post-change regime. A discounted cost criterion with a discount factor $\beta < 1$, as chosen in Bayesian and maxmin problems, may
	not always be appropriate for the application in hand. This is because such a criterion penalizes under-performing under 
	$\mathbf{M}_0$ more than it penalizes under-performing under $\mathbf{M}_1$. In reality, we may be interested in maintaining 
	good performance under every regime. 
	
	\subsection{Oracle Policy}
	Let $\mathbf{V}_0(x)$ and $\mathbf{V}_1(x)$ be the cost/reward to go at state $x$ for the MDP models $\mathbf{M}_0$ and 
	$\mathbf{M}_1$, respectively. Suppose the following conditions are satisfied:
	\begin{enumerate}
		\item the change point is exactly known,
		\item the change point $\gamma >> 1$, 
		\item for some small $\delta > 0$, the function $|\mathbf{V}_1(x)-\mathbf{V}_1(y) | \leq \delta$, $\forall x,y$, that 
		is the reward to go for the model $\mathbf{M}_1$ is not sensitive to the initial state $x$. 
	\end{enumerate}
	The implication of the first two conditions is that it is approximately optimal to use the optimal policy for $\mathbf{M}_0$ 
	before the change point. 
	The implication of the last condition is that, no matter which policy is used for model $\mathbf{M}_0$, it is still approximately optimal to use the optimal policy for model $\mathbf{M}_1$ after the change point. 
	As a result, if all three of these conditions are satisfied, then 
	one can just implement the optimal policy for the model $\mathbf{M}_0$ before change and use 
	the optimal policy for the model $\mathbf{M}_1$ after change, 
	and achieve close to maximum average rewards. In the following, such a policy is referred to as the \textit{Oracle}. 
	
	Since the change point is unknown, it is clear that if the change can
	be reliably detected, then one can hope to achieve rewards close to what the Oracle can achieve.  
	There are powerful algorithms in the literature that one can use to detect this change as quickly as possible. 
	We briefly discuss relevant ideas and algorithms in the next section, which serves as a background to readers not familiar with the literature on quickest change detection. 
	
	\section{Background on Quickest Change Detection}\label{sec:QCD}
	In the quickest change detection literature, algorithms are developed that allows one to quickly detect a change in the distribution of a stochastic process from one law to another. The optimality properties of the algorithms are studied under various modeling assumptions and problem formulations.  The theoretical foundations were laid by the work of Wald \cite{wald1973sequential}
and Shiryaev \cite{shiryaev1963optimum}; see also \cite{veeravalli2013quickest} for a survey of the area. We discuss three such algorithms in the context of problems \eqref{eq:Prob} and \eqref{eq:ProbNonBayes}. 
	
	Suppose we employ a certain policy $\Pi$ and observe a sequence of states and actions (driven by the policy) $\{(s_k, a_k)\}$. At the change point $\Gamma$, modeled as a random variable with prior probability mass function $\phi$, the model changes from $\mathbf{M}_0$
	to $\mathbf{M}_1$. Thus, the law of the stochastic sequence $\{(s_k, a_k)\}$ changes. 
	Then the following algorithm, called the Shiryaev algorithm \cite{tartakovsky2005general}, 
	has strong optimality properties. The Shiryaev algorithm 
	dictates
	that we compute the following statistic at each time $n$:
	\begin{equation}\label{eq:Shiryaevstat}
		\begin{split}
			S_n = \sum_{1 \leq k \leq n} \phi(k) \prod_{i=k}^n \frac{\mathcal{T}_1(s_{i-1}, a_{i-1}, s_i)}{\mathcal{T}_0(s_{i-1}, a_{i-1}, s_i)} ,
		\end{split}
	\end{equation}
	and declare that a change in the model has occurred at the stopping time
	\begin{equation}\label{eq:ShiryaevStop}
		\begin{split}
			\tau_{s} = \min\{ n \geq 1: S_n > A\} .
		\end{split}
	\end{equation}
	Here, $A$ is a threshold chosen to control false alarms. If the prior on the change point is geometric with 
	parameter $\rho$, then the statistic $S_n$ can be computed recursively:
	\begin{equation}\label{eq:ShiryaevRecur}
		\begin{split}
			S_n = \frac{1 + S_{n-1}}{1-\rho} \frac{\mathcal{T}_1(s_{n-1}, a_{n-1}, s_n)}{\mathcal{T}_0(s_{n-1}, a_{n-1}, s_n)}; \quad S_0 = 0.
		\end{split}
	\end{equation}
	It is shown in \cite{tartakovsky2005general} that if $A = \frac{1-\zeta}{\zeta}$ then
	\begin{equation}\label{eq:ShirOpt}
		\Expect[(\tau_s-\Gamma)^+] \sim \min_{\tau: \Prob(\tau \leq \Gamma) \leq \zeta} \Expect[(\tau -\Gamma)^+]  \sim \frac{|\log \zeta |}{I_\pi + d} \mbox{ as } \zeta \to 0. 
	\end{equation}
	Thus, the Shiryaev algorithm achieves the best average detection delay over all stopping times that satisfied a given probability of false alarm constraint of $\zeta$, as $\zeta \to 0$. Here, $d$ is a constant that is a function of the prior $\phi$ and equals $|\log(1-\rho)|$ for $\Gamma \sim$ Geom($\rho$).  
	The quantity $I_\pi$ is the Kullback-Leibler information number defined by
	\begin{equation}\label{eq:KLInfo}
		I_\pi = \lim_{n\to \infty} \frac{1}{n} \sum_{k=1}^n \log \left( \frac{\mathcal{T}_1(s_{k-1}, a_{k-1}, s_k)}{\mathcal{T}_0(s_{k-1}, a_{k-1}, s_k)}\right) \mbox{ under } \; \Prob_1. 
	\end{equation}
	Thus, larger this number $I_\pi$ is, the smaller is the detection delay. We use the subscript $\pi$ to emphasize that 
	the information number is a function of the policy $\Pi$ employed. 
	
	If the prior on the change point is not known, or if a prior cannot be defined, one can replace the Shiryaev statistic by the 
	CUmulative SUM statistic (CUSUM) \cite{lai1998information}
	\begin{equation}\label{eq:CUSUMRecur}
		W_n = \max_{n-m \leq k \leq n} \sum_{i=k}^n \log \frac{\mathcal{T}_1(s_{i-1}, a_{i-1}, s_i)}{\mathcal{T}_0(s_{i-1}, a_{i-1}, s_i)},
	\end{equation}
	where $m$ is a window size that depends on the false alarm constraint. 
	The CUSUM algorithm has strong optimality properties, similar to \eqref{eq:ShirOpt}, 
	but with average delay and probability of false alarm expressions replaced by suitable minimax delay and mean time to false alarm expressions \cite{lai1998information}. Specifically, if 
	\begin{equation}\label{eq:CusumStopRule}
		\tau_{c} = \min\{ n \geq 1: W_n > A\} ,
	\end{equation}
	then, with $A=\log \eta$,
	\begin{equation}\label{eq:ShirRobOpt}
		\begin{split}
			\sup_\gamma \; \Expect_\gamma \; [\tau_{c}-\gamma | \tau_{c}>\gamma] &\sim \min_{\tau: \Expect_\infty[\tau] \geq \eta} \sup_\gamma \; \Expect_\gamma \; [\tau-\gamma | \tau>\gamma]  \\
			& \sim \frac{|\log \eta |}{I_\pi} \mbox{ as } \eta \to \infty. 
		\end{split}
	\end{equation}
	Here, $\Expect_\gamma$ is the expectation when the change point is at $\gamma$, and $\Expect_\infty$ is
	the expectation when the change never occurs. Thus, the CUSUM algorithm minimizes the maximum of detection 
	delay--maximum over all possible change points--subject to a constraint $\eta$ on the mean time to false alarm, as 
	$\eta \to \infty$. 
	
	Again, 
	the optimal performance, for a given false alarm, depends on the information number $I_\pi $. 
	Thus, the number $I_\pi$ is the fundamental quantity 
	of interest in quickest change detection problems. It will also play a fundamental role in the rest of the paper. 
	
	An alternative to CUSUM in the non-Bayesian setting is to put $\rho=0$ in the Shiryaev algorithm and obtain
	what is called the Shiryaev-Roberts (SR) statistic
	\begin{equation}\label{eq:ShiryaevRobRecur}
		\begin{split}
			SR_n = (1 + SR_{n-1}) \frac{\mathcal{T}_1(s_{n-1}, a_{n-1}, s_n)}{\mathcal{T}_0(s_{n-1}, a_{n-1}, s_n)}; \quad SR_0 = 0,
		\end{split}
	\end{equation}
	and use the stopping time 
	\begin{equation}\label{eq:SRStopRule}
		\tau_{sr} = \min\{ n \geq 1: SR_n > A\} .
	\end{equation}
	In the numerical results reported in Section~\ref{sec:numerical}, we use the SR test instead of CUSUM because of the ease of implementation. 
	
	The algorithms discussed above can be easily modified to also account 
	for a change of distribution of rewards. We do not discuss that here for simplicity of exposition. Also, the state and action 
	sequences are more informative. Thus, unless only the reward undergoes a change, and the transition functions do not change, 
	only then it will be useful to consider rewards for detection. 
	
	\subsection*{Trade-off Between Detection Performance and Reward}
	Note that the problems in \eqref{eq:Prob} and \eqref{eq:ProbNonBayes} 
	are not classical change detection problems. The objective is not to optimize delay subject to constraint on false alarms, 
	but to optimize long term rewards. Thus, while quick detection is key, what is optimal for change detection may not be optimal with respect to optimizing rewards, and vice versa. 
	Thus, there is a fundamental trade-off between detection performance 
	and maximizing rewards. See Section~\ref{sec:QCD_Local} below for a more rigorous statement. 
	
	The fundamental contribution of this paper is a computationally efficient two-threshold algorithm that optimizes this detection-reward trade-off to achieve near optimal performance. As discussed in the introduction, most of the works in the existing literature either do not exploit tools from the change detection literature, or very few who exploit them, ignore this trade-off in their design. 
	
	%
	%Below, we suggest two such possibilities for the policy $\Pi$, 
	%and compare them with the optimal solution from the POMDP formulation of \eqref{eq:Prob}. 
	
	%For simplicity of exposition we assume that we only have two models $\mathbf{M}_0$ and $\mathbf{M}_1$, and the 
	%underlying model switches from $\mathbf{M}_0$ and $\mathbf{M}_1$ at time $\gamma$. 
	
	\section{Quickest Detection with Locally Optimal Solution}\label{sec:QCD_Local}
	One approach to solving either \eqref{eq:Prob} or \eqref{eq:ProbNonBayes} is to use a locally optimal approach, 
	i.e., to start with 
	using the optimal policy for model $\mathbf{M}_0$, compute the change detection statistics 
	(Shiryaev, SR or CUSUM) over time, 
	and switch to the optimal policy for the model $\mathbf{M}_1$ at the time when the change detection algorithm crosses its threshold.  
	Mathematically, let $\Pi_0=(\pi_0, \pi_0, \cdots)$ be the (stationary) Markov optimal policy for model $\mathbf{M}_0$, and let 
	$\Pi_1 = (\pi_1, \pi_1, \cdots)$ be the (stationary) Markov optimal policy for model $\mathbf{M}_1$. Then use the following policy
	\begin{equation}\label{eq:Policy_loc}
		\Pi_{\text{loc}} = ( \underbrace{\pi_0, \pi_0, \cdots, \pi_0}_{\tau -1 \text{ times}}, \underbrace{\pi_1, \pi_1, \cdots}_{\tau \text{ onward}} ),
	\end{equation}
	where $\tau$ is the stopping time for the Shiryaev, SR or the CUSUM algorithm. 
	
	\paragraph{Choosing the threshold} In the quickest change detection problem, the threshold $A$ used in the Shiryaev, SR or CUSUM tests is designed to satisfy the constraint on the rate of false alarms. However, for problems in \eqref{eq:Prob}
	and \eqref{eq:ProbNonBayes}, there is no notion of a false alarm. For the non-Bayesian criterion, one can choose 
	the threshold to satisfy the constraint of $\alpha$ (see \eqref{eq:ProbNonBayes}). However, in the Bayesian setting, the threshold is a free parameter, 
	and can be optimized for optimal performance. 
	
	\paragraph{Issues with policy $\Pi_{\text{loc}}$}
	As we will show in Section~\ref{sec:numerical}, the policy $\Pi_{\text{loc}}$ may lead to poor performance. This is because, 
	as discussed in Section~\ref{sec:QCD}, the detection performance of the stopping rule $\tau$ used depends on 
	the information number $I_{\pi_0}$; see \eqref{eq:KLInfo}. Let 
	\[
	I_{\max} \; := \; \max_\pi \; I_\pi. 
	\]
	Then, it is possible that 
	\[
	I_{\pi_0}\; \ll I_{\max},
	\]
	and that $I_{\pi_0}$ is itself quite small. This may lead to poor detection performance, and hence may cause significant 
	loss of revenue or rewards. If it so happens that $I_{\pi_0}\; \approx I_{\max}$, and $I_{\pi_0}$ is significant, then 
	one can expect that this simple strategy $\Pi_{\text{loc}}$ itself may lead to near optimal performance. 
	But, this is more a matter of chance, and  more sophisticated approaches are needed that considers every possibility. 
	We note that in some cases one may be able to use MOMDP or POMDP based approximations to compute the optimal policy
	for our problem, but such solutions are computationally demanding. Using a quickest change detection approach leads
	to computationally efficient algorithmic techniques. 
	
	\section{A Two-Threshold Switching Strategy to Exploit Reward-Detection Trade-Off}\label{sec:QCD_Twothres}
	One way to improve the performance of the policy $\Pi_{\text{loc}}$ is to replace
	the policy $\Pi_0$ by another policy that better exploits the reward-detection trade-off.  
	One option is to replace $\Pi_0$ by a policy that maximizes the information number $I_\pi$. 
	Define for each state $s \in \mathcal{S}$, the information maximizing policy $\pi_{\text{KL}}(s)$ as 
	\begin{equation}
		\pi_{\text{KL}}(s) = \text{ arg max}_a \sum_{s'} \mathcal{T}_1 (s, a, s') \; \log \frac{\mathcal{T}_1 (s, a, s')}{\mathcal{T}_0 (s, a, s')}.
	\end{equation}
	Specifically, use
	\begin{equation}\label{eq:Policy_KL}
		\Pi_{\text{KL}} = ( \underbrace{\pi_{\text{KL}}, \pi_{\text{KL}}, \cdots, \pi_{\text{KL}}}_{\tau -1 \text{ times}}, \underbrace{\pi_1, \pi_1, \cdots}_{\tau \text{ onward}} ).
	\end{equation}
	Using this policy will lead to quickest and most efficient detection of model changes. However, since the policy $\pi_\text{KL}$ may not be optimal for the MDP model $\mathbf{M}_0$, it may cause significant loss of rewards before the change point. 
	
	We propose to use a simple two-threshold strategy that switches between $\pi_0$ and $\pi_{\text{KL}}$, using $\pi_0$ to 
	optimize rewards, and switching to $\pi_{\text{KL}}$ for information extraction. The proposed switching strategy 
	is closely related to the notion of sequential design of experiments \cite{chernoff1959sequential},
	or to the notion of exploitation and exploration in multi-arm bandit problems~\cite{gittins2011multi}. 
	
	The two-threshold policy $\Pi_{\text{TT}}$ is described in Algorithm~\ref{algo:Algorithm}. In words, fix two thresholds $A$ and $B$, with $B<A$. Also, assume that we are in the Bayesian setting so that we can compute the Shiryaev statistic $S_n$  \eqref{eq:ShiryaevRecur}. Then the policy $\Pi_{\text{TT}}$ is defined as follows. 
	Start with the locally optimal policy $\Pi_0=(\pi_0, \pi_0, \cdots)$ and compute the statistic $S_n$. As long as $S_n \leq B$, keep using the Markov policy $\pi_0$. 
	If $S_n > B$, suggesting that we may be in the wrong model, extract more information by using
	$\pi_{\text{KL}}$.  If $S_n$ goes below $B$ again, switch back to $\pi_0$, else we keep using $\pi_{\text{KL}}$. If $S_n > A$, switch to the optimal Markov policy for model $\mathbf{M}_1$, that is $\pi_1$. 
	\begin{figure}[t]
		\begin{algorithm}{$\Pi_{\text{TT}}$: Two-Threshold Switching Strategy}\label{algo:Algorithm}
			%	\captionsetup{font=normalsize}
			%\caption{\textsc{Two-Threshold Switching Strategy}}
			\algrule
			\begin{algorithmic}[1]
			%	\algrule\\
				\Require Thresholds $A$ and $B$, $B<A$; optimal policies $\pi_0$, $\pi_1$, $\pi_{\text{KL}}$; transition kernels $\mathcal{T}_1(s,a,s')$
				and $\mathcal{T}_0(s,a,s')$
				\State Start with $S_0=0$%Construct a validity map $\mathcal{G}_\Psi$
				\While{$S_n \leq A$}
				\State Use locally optimal policy $\Pi_0=(\pi_0, \pi_0, \cdots)$
				\State Compute the statistic $S_n$ using \eqref{eq:ShiryaevRecur}
				\State If $S_n \leq B$ continue using the policy $\pi_0$
				\State If $S_n > B$ use $\pi_{\text{KL}}$
				\State n = n + 1
				\EndWhile
				\State Switch to policy $\Pi_1=(\pi_1, \pi_1, \cdots)$
				\State \Return %Controller parameters, $\{\Theta_n\}_{n=1}^{|N|}$. %and the posterior of distribution of latent variables $\phi$.
				\algrule
			\end{algorithmic}
		\end{algorithm}
		%\vskip -0.2in
	\end{figure}
	
	In short, use $\pi_0$ when $S_n \in [0, B] $, use $\pi_{\text{KL}}$ when $S_n \in (B, A]$, and use $\pi_1$ when  
	$S_n \in (A, \infty)$. Thus, the change detection statistic $S_n$ is used as a belief on the unknown model, and is used for detection between $\mathbf{M}_0$ and $\mathbf{M}_1$, depending on whether $S_n$ is small or large, respectively. For moderate values of $S_n$, the policy $\pi_{\text{KL}}$ is used to improve detection performance.  
	\begin{equation}\label{eq:Policy_TT}
		\begin{split}
			\Pi_{\text{TT}} &= ( \underbrace{\pi_0, \cdots, \pi_0}_{S_n \leq B}, 
			\underbrace{\pi_{\text{KL}},  \cdots, \pi_{\text{KL}}}_{B < S_n \leq A}, 
			\underbrace{\pi_0, \cdots, \pi_0}_{S_n \leq B}, \cdots, \\
			&\hspace{2cm} \cdots, \underbrace{\pi_{\text{KL}},  \cdots, \pi_{\text{KL}}}_{B < S_n \leq A \text{ until } \tau -1}, \underbrace{\pi_1, \pi_1, \cdots}_{\tau \text{ onward}} ).
		\end{split}
	\end{equation}
	In the non-Bayesian setting, we can replace the Shiryaev statistic $S_n$ by the SR statistic $SR_n$ \eqref{eq:ShiryaevRobRecur} or the CUSUM statistic $W_n$ \eqref{eq:CUSUMRecur}. 
	
	\paragraph{On the choice of thresholds $A$ and $B$} Note first that setting $A=B$ reduces the two-threshold policy
	$\Pi_{\text{TT}}$ to the locally optimum policy $\Pi_{\text{loc}}$. Also, by setting $B=0$ the policy reduces to $\Pi_{\text{KL}}$. 
	As discussed earlier, this will lead to good detection performance, but may lead to loss of rewards. 
	Akin to $\Pi_{\text{loc}}$, the thresholds $A$ and $B$ here
	are free parameters, even for the non-Bayesian setting. Thus, the thresholds can be optimized for best performance: choosing
	the optimal thresholds is equivalent to choosing the best detection-reward trade-off. 
	
	\paragraph{Bayesian vs Non-Bayesian} We emphasize that the definition of $\Pi_{\text{TT}}$ is implicitly different for Bayesian and non-Bayesian cases. In the Bayesian case, we have access to the prior on the change point, and we use that to compute the Shiryaev 
	statistic $S_n$. However, in the non-Bayesian case, we replace the Shiryaev statistic by the Shiryaev-Roberts statistic $SR_n$ or the CUSUM statistic $W_n$. 
	
	\paragraph{When will $\Pi_{\text{TT}}$ outperform $\Pi_{\text{loc}}$?} The amount of performance gain depends in general 
	on the problem structure: reward function and the transition kernel. 
	However, a condition which must be satisfied is $I_{\pi_0} \ll I_{\max}$. This condition often leads
	to  better detection performance for $\Pi_{\text{TT}}$ as compared to $\Pi_{\text{loc}}$, resulting in efficient detection and more average rewards. 
	Another condition under which performance gains are significant is when
	under-performing under any of the models is penalized equally. 
	For example, in the Bayesian case \eqref{eq:Prob}, if the discount factor $\beta$ is small, then the rewards lost in 
	delaying a transition to the optimal policy of $\mathbf{M}_1$ will not affect the overall cost. As a result, 
	a policy can cause large delays, without significantly loosing rewards, and a good detection performance may not 
	necessarily lead to noticeable gain in rewards. 
	Numerical results are provided in the next section to support these observations.

	%\textbf{Consistency}: Under mild conditions, the policy $\Pi_{\text{TT}}$ can be proved to reliably detect the change
	%see \cite{tartakovsky2005general} and \cite{lai1998information} for detailed discussion. We state the result here for the 
	%Bayesian setting when we use the Shiryaev statistic. 
	%\begin{theorem}[\cite{tartakovsky2005general}]
	%Let 
	%\[
	%\lim_{n\to \infty} \frac{1}{n} \sum_{k=1}^n \log \left( \frac{T_1(s_{k-1}, a_{k-1}, s_k)}{T_0(s_{k-1}, a_{k-1}, s_k)}\right) \to I^* > 0 \mbox{ a.s. under } \; \Prob_1, \; \mbox{ and policy } \Pi_{\text{TT}}, 
	%\]
	%and the notion of average-r-quick convergence is satisfied (see Theorem 2, \cite{tartakovsky2005general}), then the Shiryaev
	%test applied to the state-action sequence generated by $\Pi_{\text{TT}}$ minimizes all moments of the detection delay, as the %probability of false alarm goes to zero. 
	%\end{theorem}
	
\iftrue
	\section{Multiple Models and Change Points}\label{sec:multModel_multCP}
%	In practice, there is more than one way in which the environment can change. 
%	The restricted assumption of two possible models, one pre- and another post-change, 
%	was made only to simplify notation and for ease of exposition. The ideas developed
%	can easily be extended to multiple, even infinite number of models. 
	
	For a parametrized family of models $\{\mathbf{M}_\theta\}$, if the model changes 
	from $\mathbf{M}_{\theta_0}$ to some other unknown model in the family, then the Shiryaev, SR or CUSUM statistic
	cannot be used for change detection. This is because one needs the exact knowledge of the post-change 
	distribution to compute those statistics. Popular alternatives are to use generalized likelihood ratio (GLR) 
	based tests or mixture based tests. We only discuss the former. A GLR statistic for change detection is defined as
	\[
	G_n = \max_{n-m \leq k \leq n} \sup_{\|\theta-\theta_0\| \geq \epsilon} \; \sum_{i=k}^n \log \frac{\mathcal{T}_\theta(s_{i-1}, a_{i-1}, s_i)}{\mathcal{T}_{\theta_0}(s_{i-1}, a_{i-1}, s_i)},
	\]
	where $\epsilon$ is the minimum magnitude of change that can occur in the problem. 
	
	A replacement for $\Pi_{\text{loc}}$ in this setting would be to use locally optimal policy and switch at the time of stopping, 
	where now the stopping rule is 
	\[
	\tau_g = \min\{ n \geq 1: G_n > A\}.
	\]
	Since there is more than one possibility for the post-change model, one can use the maximum likelihood estimate (the $\theta$ 
	that achieved the maximum in the expression for $G_n$ at the time of stopping) to choose the model. When the number of models is finite, one can 
	also use the theory of fault isolation to choose the right model \cite{tartakovsky2008multidecision}. 
	
	A replacement for $\Pi_{\text{TT}}$ in this setting is even less obvious because of the difficulty in defining an appropriate 
	version of $\pi_{\text{KL}}$. We propose the following
	\begin{equation}
		\pi_{\text{KL}}(s) = \text{ arg max}_a \min_{\|\theta-\theta_0\| \geq \epsilon} \sum_{s'} \mathcal{T}_\theta (s, a, s') \; \log \frac{\mathcal{T}_\theta (s, a, s')}{\mathcal{T}_{\theta_0} (s, a, s')}.
	\end{equation}
	Thus, maximize the worst case KL divergence. Analytical and numerical performance of these schemes will be reported elsewhere. 
	
	If there are multiple change points, then as long as the gap between the change points is large enough, we can repeat 
	the two model algorithm for consecutive change points. Specifically, if the GLR based algorithm detects a change 
	and gives $\theta_1$ as an estimate for the post-change parameter, then we can reset the algorithm with $\theta_1$ as the 
	pre-change parameter, and reapply the GLR based test. As long as there is a significant gap between model parameters, 
	one would expect this technique to be close to optimal. 
	
%	In this paper we assumed that the transition kernel family $\mathcal{T}_\theta$ is parametric, and is known to the decision maker. 
	%It is possible that in practice such an assumption is restrictive. In such a case, we believe that techniques from the 
	%reinforcement learning literature \cite{kaelbling1996reinforcement} can be combined with the ideas from this paper to develop efficient algorithms, which is a topic of future research.
%	We do not discuss such approaches in this paper. 
\fi

	\section{Numerical Results}\label{sec:numerical}
	We now compare the two policies $\Pi_{\text{TT}}$ and $\Pi_{\text{loc}}$ for an inventory control problem, 
	and show that using $\Pi_{\text{TT}}$ results in significant gains. We use the inventory control problem 
	to illustrate the ideas because of its simplicity. The understanding developed through this simple problem 
	can be extended to more complex problems. 
	
	\paragraph{Inventory control problem} 
		The state $s_k$ is the level of the inventory at time $k$, 
	and the maximum size of the inventory is $N$. The action $a_k$ 
	is the additional inventory to be ordered based on the state $s_k$, making the total inventory $s_k+a_k$. 
	A stochastic demand $w_k$ arrives making the residual inventory or the next state $s_{k+1} = \max\{0, s_k + a_k - w_k\}$.  
	Let $c$ to be cost of ordering a unit of inventory, $h$ be the holding cost per unit, and $p$ be the cost of loosing a unit of demand.
	Thus the cost per unit time is given by
	\begin{equation}
		\begin{split}
			C(s_k, a_k) &= \Expect_{w_k} \left[\; c a_k + h \max\{0, s_k + a_k - w_k\}  \right.\\
			& \hspace{1cm} \left. + p \max\{0, -(s_k + a_k - w_k)\} \right].
		\end{split}
	\end{equation}  
	It is assumed that at the change point, the distribution of the demands changes from Poisson($\lambda$) to Uniform. 
	The demands are assumed to be independent, conditioned on the change point. 
For clarification, note that the problem formulations \eqref{eq:Prob} and \eqref{eq:ProbNonBayes}
are about maximizing rewards, but the objective in the inventory control problem is to minimize cost. 	
	
	\paragraph{Numerical results for Bayesian setting}
	Under the Bayesian setting \eqref{eq:Prob}, it is assumed that the change occurs with a geometric prior $\rho=0.01$.
	The other parameter values used are $c=1$, $h=5$, $\beta=0.99$, $\lambda=2$, $N=10, 20$, and $p$ is chosen to be $100, 200$ or $300$.  
	This is a classical optimal control problem and the optimal policy under each model, Poisson or Uniform, 
	can be found using value iteration \cite{bertsekas1995dynamic}. 
	The performance of various policies discussed are tabulated in Table~\ref{Tab:Bayesian}. 
	The results are obtained by performing average over $1000$ independent runs, each of horizon size $1000$. The Oracle policy
	is the one that knows the change point, and switches from the optimal policy for the Poisson model to the optimal policy 
	for the Uniform model, exactly at the change point. The policy $\Pi_{\text{loc}}$ employs the Shiryaev algorithm \eqref{eq:ShiryaevStop}. Since there is no concept of false alarms, the choice of threshold $A$ is optimized to achieve
	minimum possible cost. The policy $\Pi_{\text{TT}}$ also employs the Shiryaev algorithm, and the values of the two 
	thresholds are also chosen to achieve minimum possible cost. For comparison, the costs achieved by performing
	an MOMDP approximation and a random action strategy is also shown. The performance of the policy $\Pi_{\text{TT}}$ is comparable 
	to that of the MOMDP solution, and significantly better than that of $\Pi_{\text{loc}}$. Note that the gain is more significant 
	when $I_{\max}$ is much larger than $I_{\pi_0}$. 
	Also, since the Poisson arrival rate chosen implies less demand as compared to the Uniform arrival rate, the optimal control is more aggressive for the Uniform arrival model. A delay in detecting the change will result in significant loss of demands. 
	The chosen value of the discount factor $\beta$ and $p$ ensures 
	that delays in detection, and hence loss in demands, are sufficiently penalized under both regimes, pre-change and post-change. 
	\begin{table} [t]
		\centering
		%\hspace{1.7cm}
		%\begin{minipage}[t]{0.7\linewidth}
		\centering
		{\scriptsize
			\caption{\small Bayes discounted cost \eqref{eq:Prob} for different policies: $c=1$, $h=5$, $\beta=0.99$, $\rho=0.01$. The values 
				in the table are obtained by averaging costs over $1000$ independent runs, each of horizon size $1000$.}
			\label{Tab:Bayesian}
			\scalebox{0.95}{
				\begin{tabular}{|l|l|l|l|l|l|l|l|l|l|l|l|l|}
					\hline
					$N$&$p$ & Oracle &   $\Pi_{\text{loc}}$              &   $\Pi_{\text{TT}}$      & MOMDP   &RANDOM&$I_{\max}$ & $I_{\pi_0}$\\
					\hline
					20&100 & 4057 & 4980 &4579&4688&16489&8.45&2.56\\
					20&200 & 4229 & 5678 &5051&5144&27655&8.45&2.56\\
					20&300 & 4277 & 5884 &5238&5325&38821&8.45&3.24\\
					10 &  100 &2336 & 2619&2515&2677&7861&1.72&1.17\\
					10 &  200 & 2445&2902 &2728&2876&13623&1.72&1.42\\
					10 &  300 &2550 &2920 &2813&3020&19386&1.72&1.61\\
					\hline
				\end{tabular}}
			}
		\end{table}
		
		%\pagebreak
		
\begin{figure}[t]
			\center
			\includegraphics[width=6.5cm, height=4.5cm]{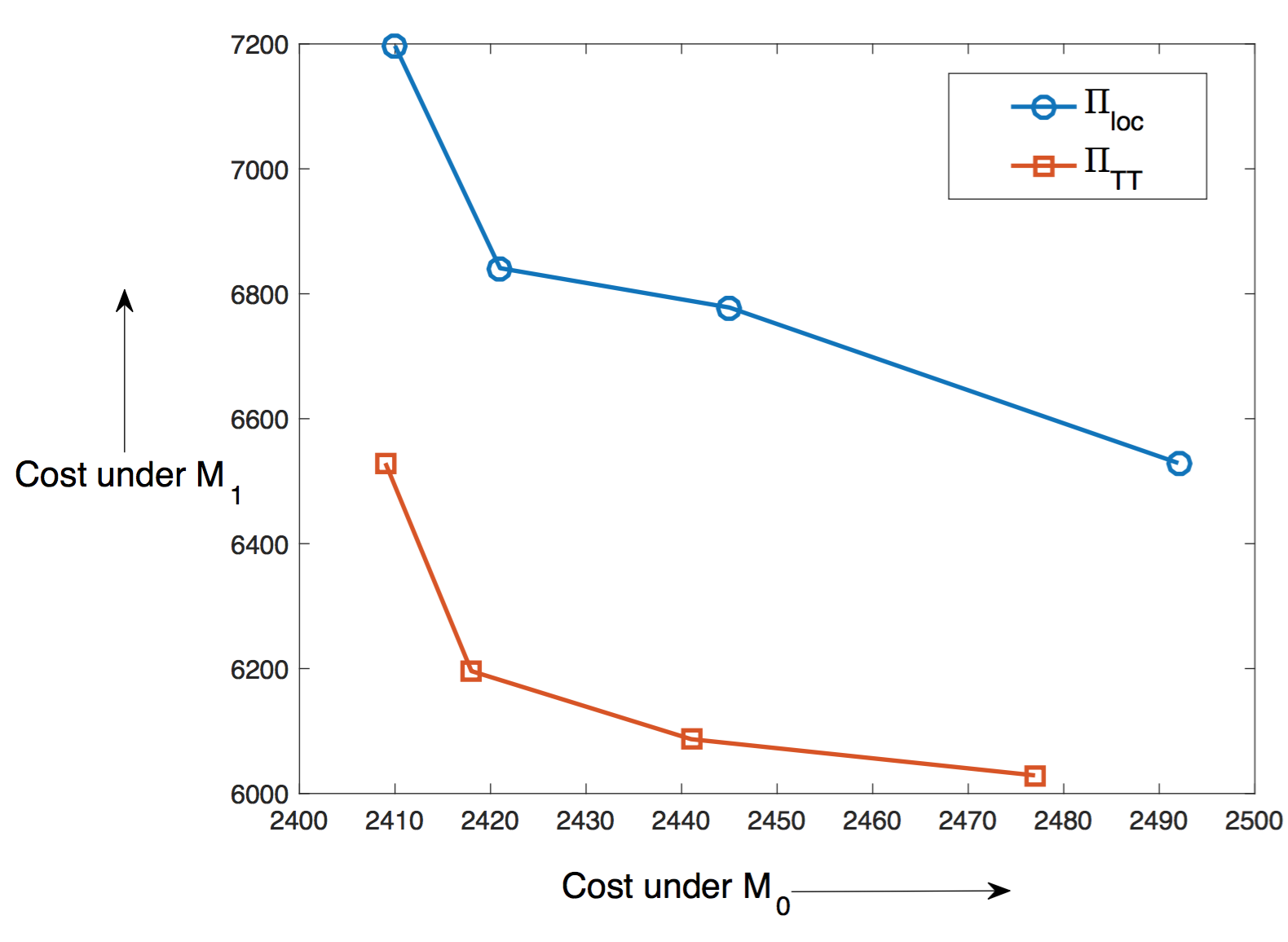}
			\caption{\small Performance under $\mathbf{M}_1$ subject to constraints on the performance under $\mathbf{M}_0$. The SR algorithm \eqref{eq:ShiryaevRobRecur} is used while implementing the policies.}
			\label{fig:NonBayes}%\vspace{-1cm}
		\end{figure}

		\paragraph{Numerical results for Non-Bayesian setting}
		An overall discounted criterion like \eqref{eq:Prob} may not be the best or even fair
		way to compare competing algorithms for a non-stationary environment. 
		One of the primary reasons for using \eqref{eq:Prob} as a criterion was to show that the proposed 
		algorithm can perform comparable to an MOMDP approximation. 
		A more appropriate performance metric is a variational one \eqref{eq:ProbNonBayes}. 
		In this setting, instead of optimizing the threshold used in the policies, the thresholds are selected to satisfy the constraint of $\alpha$ 
		on $ \Expect_\infty \left[ \sum_{k=0}^\infty \beta^k C_k (s_k, a_k)\right] $, in which the change never occurs. For these choices
		of the thresholds, the performance $ \Expect_1 \left[ \sum_{k=0}^\infty \beta^k C_k (s_k, a_k)\right] $ is evaluated for both  $\Pi_{\text{TT}}$ and $\Pi_{\text{loc}}$ in which the change occurs at time $1$. 
		This way, performance under both the regimes or MDPs are equally weighted,
		and an algorithm that performs well in every setting can be obtained. 
		The performance comparison is plotted in  
		Fig.~\ref{fig:NonBayes}. 
		The parameter used are $c=1$, $h=5$, $\beta=0.99$, $\lambda=2$, $N=20$, and $p=200$.  
		We use the Shiryaev-Roberts \eqref{eq:ShiryaevRobRecur} algorithm to detect changes.  Clearly, 
		$\Pi_{\text{TT}}$ is superior. Finally, note that since we do not have a prior on the change point, a POMDP or MOMDP solution 
		cannot be obtained. 
		%\begin{figure}[htb]
		%\center
		%\hspace{3cm}
		%\includegraphics[width=8cm, height=5cm]{NIPS_NonBayes.pdf}
		%\caption{Comparative performance under $\mathbf{M}_1$ of policies $\Pi_{\text{TT}}$ and $\Pi_{\text{loc}}$ subject
		%to constraints on the performance under $\mathbf{M}_0$.}
		%\label{fig:NonBayes}
		%\end{figure}
		
		\section{Conclusions}\label{sec:discussion}
		This paper presents a novel way to combine techniques used for stationary MDPs and quickest change detection to solve non-stationary MDPs by considering the trade-off between change detection and reward maximization, an important problem that has not been adequate addressed before. Our method uses a two threshold switching strategy to exploit reward-detection trade-off. Our numerical results show that the proposed method can achieve better trade-off, and outperform the state-of-the art QCD method and MOMDP method for solving non-stationary MDP tasks. Future work will include application of ideas developed in this paper to design of reinforcement learning based algorithms for decision making in non-stationary environments. See \cite{SinghBanerjee2017} for a stochastic approximation based approach to this problem. 
		
		\section*{Acknowledgment}
		This research was funded in part by Northrop Grumman and Lincoln Laboratory.

		\balance
		{%\footnotesize
			\bibliographystyle{IEEEtran.bst}
			\bibliography{nips_2016}
		}

	\end{document}